\documentclass{article}
\usepackage{amsmath}

\newtheorem{theorem}{Theorem}
\newtheorem{acknowledgement}[theorem]{Acknowledgement}

\input{tcilatex}

\begin{document}

\title{Local field modulated entanglment among three distant atoms}
\author{Yan-Qing Guo\thanks{%
E-mail:guoyqthp@yahoo.com}, Jing Chen, He-Shan Song\thanks{%
Corresponding author: hssong@dlut.edu.cn} \\
%EndAName
Department of Physics, Dalian University of Technology \\
Dalian, 116024, P. R. China}
\maketitle

\begin{abstract}
We extend the scheme for that proposed by S. Mancini and S. Bose (Phys. Rev.
A \textbf{70}, 022307(2004)) to the case of triple-atom. Under mean field
approximation, we obtain an effective Hamiltonian of triple-body Ising-model
interaction. Furthermore, we stress on discussing the influence of the
existence of a third-atom on the two-atom entanglement and testing the
modulation effects of locally applied optical fields and fiber on the
entanglement properties of our system.

\end{abstract}

\baselineskip=15.5pt

\section{Introduction}

Generating and maintaining the entanglement between two spatially seperated
atoms plays an important role in many quantum information processing, such
as quantum storage\cite{1}, quantum key distribution\cite{2} and quantum
states swapping\cite{3}, since need no transmitting real particles to
distance. To efficiently entangle two distant atoms, one must creat some
kind of interaction that is off-diagonal in its energy eigenstates
representation. Generally, such interaction is equivalent to exchange real
photons between two spatially related atoms. Many schemes have been proposed
to realize this kind of entanglement[4-13]. Most of these schemes involve
two cavities in each of which an atom is trapped. In Ref. [4], the author
proposed such a scheme. He showed how an effective direct interaction
between two atoms is established. The author used photon as an intermediate
quantum information carrier to map the quantum information from an atom in
one cavity to an atom in another cavity. By eliminating the optical fields
ingoing and outgoing the local cavity, the author found that two distant
atoms can directly interact in a form of Ising spin-spin coupling. It is
meaningful conceptually in quantum measurement and testing Bell's
inequalities and engineeringly in quantum encryption\cite{14} and
constructing universal quantum gate\cite{15} that are essential elements for
designing quantum network. While, in realizing quantum network, an atomic
ensemble is usually required\cite{1}. The existence of collective
interaction inevitably influence the entanglement shared by atoms. In this
paper, we extend the model in Ref. [4] to triple-atom situation. We focus on
investigating the affect of the existence of the third atom on the
entanglement properties and the modulation of operating local optical field
plus fiber on the information signal. This system can be easily extended to
N-atom system.

\section{Optical Fiber Connected Three-Atom-Cavity System}

Fig. 1 shows the imagined triple-atom composed setup. Two-level atoms 
\textit{A}, \textit{B} and \textit{C} locate in spatially distant cavities
1, 2 and 3 respectively. Cavities 1 and 3 are assumed to be single-sided
cavities, cavity 2 to be double-sided cavity. An off-resonant driving
external field A is added on cavity 1. In each cavity, a local laser field
which is resonantly coupled to the local atom is applied. Two neighbouring
cavities are connected with optical fiber.

The Hamiltonian for the global system can be written as

\begin{equation}
H=H_{atom}+H_{field}+H_{bath}+H_{int1}+H_{int2}\text{.}
\end{equation}

In the interaction picture, The Hamiltonian should include the interactions
between atom and bath and that between field and bath

\begin{equation}
H_{int}=H_{in1}+H_{int2}\text{,}
\end{equation}%
where\cite{16}

\begin{equation}
H_{int1}=\chi A^{\dagger }A\sigma _{1}^{z}+\chi B^{\dagger }B\sigma
_{2}^{z}+\chi C^{\dagger }C\sigma _{3}^{z}\text{,}
\end{equation}%
and\cite{17}

\begin{eqnarray}
H_{int2} &=&i\int\limits_{-\infty }^{+\infty }d\omega \{\kappa _{A}\left[
b_{A}(\omega )A^{\dagger }+h.c.\right] +\kappa _{B,L}\left[ b_{B,L}(\omega
)B^{\dagger }+h.c.\right]  \notag \\
&&\text{ \ \ \ }+\kappa _{B,R}\left[ b_{B,R}(\omega )B^{\dagger }+h.c.\right]
+\kappa _{C}\left[ b_{C}(\omega )C^{\dagger }+h.c.\right] \}\text{,}
\end{eqnarray}
where $A(A^{\dagger })$, $B(B^{\dagger })$ and $C(C^{\dagger })$ the field
annihilation (creation) operators in cavities A, B and C, respectively, $%
\chi =g^{2}/\Delta $ with $g$ the dipole coupling strength of atom on field
and $\Delta $ the detuning of field from the atomic internal transition, $%
\sigma _{i}^{z}(i=1,2,3)$ represents the particle inversion number of atom $%
i $.

The kinetic equations for the field operators turn out to be\cite{17}

\begin{eqnarray}
\dot{A} &=&-i\Delta A-i\chi A\sigma _{1}^{z}-\frac{\gamma _{A}}{2}A+\sqrt{%
\gamma _{A}}A_{in}+\text{\textrm{A},}  \notag \\
\dot{B} &=&-i\Delta B-i\chi B\sigma _{2z}-\frac{\gamma _{B,L}+\gamma _{B,R}}{%
2}B+\sqrt{\gamma _{B,L}}B_{in,L}+\sqrt{\gamma _{B,R}}B_{in,R}\text{,}  \notag
\\
\dot{C} &=&-i\Delta C-i\chi C\sigma _{2z}-\frac{\gamma _{C}}{2}C+\sqrt{%
\gamma _{C}}C_{in}\text{,}
\end{eqnarray}%
where $\gamma _{i}=2\pi k_{i}^{2}(\omega )$, \textrm{A }is the amplitude of
the input off-resonant driving field in cavity $A$.

\ If cavities $A$, $B$ and cavities $B$, $C$ are connected by fiber (as
shown in Figure 1), the input-output condition should be included, such that%
\cite{18}

\begin{eqnarray}
\dot{A} &=&-\frac{\gamma _{A}}{2}A+\sqrt{\gamma _{A}}B_{out,L}e^{i\phi _{12}}%
\text{,}  \notag \\
\dot{B} &=&-\frac{\gamma _{B,L}}{2}B+\sqrt{\gamma _{B,L}}A_{out}e^{i\phi
_{21}}\text{,}  \notag \\
\dot{B} &=&-\frac{\gamma _{B,R}}{2}B+\sqrt{\gamma _{B,R}}C_{out}e^{i\phi
_{23}}\text{,}  \notag \\
\dot{C} &=&-\frac{\gamma _{C}}{2}C+\sqrt{\gamma _{C}}B_{out,R}e^{i\phi _{32}}%
\text{.}
\end{eqnarray}
Assuming all the decay rates $\gamma _{A}=\gamma _{B,L}=\gamma _{B,R}=\gamma
_{C}=\gamma _{0}$ and taking into account the usual boundary conditions\cite%
{17}

\begin{equation}
A_{out}(B_{out,L},B_{out,R},C_{out})+A_{in}(B_{in,L},B_{in,R},C_{in})=\sqrt{%
\gamma _{0}}A(B,B,C)\text{,}
\end{equation}
we can rewrite the kinetic equations for field operators as

\begin{eqnarray}
\dot{A} &=&-(i\Delta +\gamma _{0})A-i\chi A\sigma _{1}^{z}+\sqrt{\gamma _{0}}%
A_{in}+\gamma _{0}A_{in}+e^{i\phi _{12}}(\gamma _{0}B-\sqrt{\gamma _{0}}%
B_{in,L})+\Lambda \text{,}  \notag \\
\dot{B} &=&-(i\Delta +\gamma _{0})B-i\chi B\sigma _{2}^{z}+\sqrt{\gamma _{0}}%
(B_{in,L}+B_{in,R})+e^{i\phi _{21}}(\gamma _{0}A-\sqrt{\gamma _{0}}A_{in}) 
\notag \\
&&+e^{i\phi _{23}}(\gamma _{0}C-\sqrt{\gamma _{0}}C_{in})\text{,}  \notag \\
\dot{C} &=&-(i\Delta +\gamma _{0})C-i\chi C\sigma _{3}^{z}+\sqrt{\gamma _{0}}%
C_{in}+e^{i\phi _{32}}(\gamma _{0}B-\sqrt{\gamma _{0}}B_{in,R})\text{.}
\end{eqnarray}

The phase factors $\phi _{12}$, $\phi _{21}$, $\phi _{23}$ and $\phi _{32}$
are the phase delay caused from the photon transmission along the optical
fiber. Physically, they depend on the frequency of the photons and the
distance between cavities. The upper equations are non-linear ones since
there exist cross terms include field operators multiplying atomic spin
operators. To solve the kinetic equation of fields operators explicitly,
these terms must be modified. To do this, we make such assumptions: we are
interested in the stationary quantum effects of the field-atom system;
strong leakage condition for the cavity; large detuning from the atomic
internal transition (see the assumption before). Thus, only the terms
besides the cross ones have apparent large gradients. The field operators
multiplied by atomic spin operators can be approximatly replaced by their
stationary values which can be obtained through

\begin{equation}
\frac{d\left\langle A\right\rangle }{dt}=\frac{d\left\langle B\right\rangle 
}{dt}=\frac{d\left\langle C\right\rangle }{dt}=0\text{.}
\end{equation}%
We get

\begin{eqnarray}
\alpha &=&\frac{\Lambda \lbrack M(M+\gamma _{0})-\gamma _{0}^{2}e^{i(\phi
_{23}+\phi _{32})}]}{M^{2}(M+\gamma _{0})-M\gamma _{0}[e^{i(\phi _{12}+\phi
_{21})}+e^{i(\phi _{23}+\phi _{32})}]}\text{,}  \notag \\
\beta &=&\frac{M\alpha -\Lambda }{\gamma _{0}e^{i\phi _{12}}}\text{, }\gamma
=\frac{\gamma _{0}e^{i\phi _{32}}\beta }{M}\text{.}
\end{eqnarray}

Now, the nonlinear differential Equ. [5] are linearized into

\begin{eqnarray}
\dot{a} &=&-(i\Delta +\gamma _{0})a-i\chi \alpha \sigma _{1}^{z}+\sqrt{%
\gamma _{0}}a_{in}+e^{i\phi _{12}}(\gamma _{0}b-\sqrt{\gamma _{0}}b_{in,L})%
\text{,}  \notag \\
\dot{b} &=&-(i\Delta +\gamma _{0})b-i\chi \beta \sigma _{2}^{z}+\sqrt{\gamma
_{0}}(b_{in,L}+b_{in,R})+e^{i\phi _{21}}(\gamma _{0}a-\sqrt{\gamma _{0}}%
a_{in})  \notag \\
&&+e^{i\phi _{23}}(\gamma _{0}c-\sqrt{\gamma _{0}}c_{in})\text{,}  \notag \\
\dot{c} &=&-(i\Delta +\gamma _{0})c-i\chi \gamma \sigma _{3}^{z}+\sqrt{%
\gamma _{0}}c_{in}+e^{i\phi _{32}}(\gamma _{0}b-\sqrt{\gamma _{0}}b_{in,R})%
\text{.}
\end{eqnarray}%
where we have replaced field operators $A$, $B$ and $C$ with $a+\alpha $, $%
b+\beta $ and $c+\gamma $. Equ. [11] can be easily solved with Laplace
transform method. In solving Equ. [11], we can adiabatically eliminate the
effect of vacuum input noise on the cavity field operators. The resulting
field operators are simply represented by linear combinations of atomic spin
operators $\sigma _{1}^{z}$, $\sigma _{2}^{z}$ and $\sigma _{3}^{z}$.
Substituting the field operators into the interaction Hamiltonian, we get
the effective Hamiltonian of the global system in the interaction picture as

\begin{equation}
H_{eff}=J_{12}\sigma _{1}^{z}\sigma _{2}^{z}+J_{23}\sigma _{2}^{z}\sigma
_{3}^{z}+J_{31}\sigma _{3}^{z}\sigma _{1}^{z}\text{,}
\end{equation}%
which is an triple-body Ising type interaction, similar with that in Ref.
[4]. We have defined

\begin{eqnarray}
J_{12} &=&\gamma _{0}\chi ^{2}\func{Im}\left\{ \alpha \beta ^{\ast }e^{i\phi
_{21}}/[M^{2}-W^{2}]\right\} \text{,}  \notag \\
J_{23} &=&\gamma _{0}\chi ^{2}\func{Im}\left\{ \alpha \beta ^{\ast }e^{i\phi
_{32}}/[M^{2}-W^{2}]\right\} \text{,}  \notag \\
J_{31} &=&\gamma _{0}\chi ^{2}\func{Im}\left\{ \gamma _{0}\gamma \alpha
^{\ast }[e^{i\phi _{23}}+e^{i\phi _{12}}]/[M(M^{2}-W^{2})]\right\} \text{,}
\end{eqnarray}%
with $M=i\Delta +\gamma _{0}$ and $W^{2}=\gamma _{0}^{2}\left[ \frac{1}{4}%
+e^{i(\phi _{21}+\phi _{12})}+e^{i(\phi _{32}+\phi _{23})}\right] $. $J_{12}$
and $J_{23}$ are the nearest-neighbour atom radiation pressure, while $%
J_{31} $ presents next-neighbour atom interaction strength. They all turn
out to be zero when cavity fields are resonant with atoms and three cavities
are spatially much close to each other.

We have neglected local self-energy terms including $\sigma _{i}^{z}$ by
choosing appropriate detuning $\Delta $ and self-interaction terms that do
not change the initial system state in deriving $H_{eff}$. Also, we
eliminated high order terms that include $\sigma _{1}^{z}\sigma
_{2}^{z}\sigma _{3}^{z}$ since the corresponding coupling strength is much
weaker than $J_{12}$, $J_{23}$ and $J_{31}$. In next section, laser fields
are employed to induce two-atom entanglement.

\section{Fiber Plus Laser Field Modulated Two-Atom Entanglement}

In Ising model, entanglement can be generated between neighbouring spin
sites if each of them is coupled with a magnetic field that do no parallel
the spin $z$\ axis\cite{19}. Note that the Hamiltonian in Equ. [12] can not
generate any entanglement because it is diagonal in the $2\otimes 2\otimes 2$
dimensions atomic subspace. It is necessary to add a laser field in each
cavity which is resonantly interacts with the local atom, so that the
additional Hamiltonian is given by

\begin{equation}
H_{add}=\Gamma \sum\limits_{i}(\sigma _{i}+\sigma _{i}^{+})\text{,}
\end{equation}%
where $\Gamma $ represents the coupling strength of the laser field and the
local atom, $i$ denotes atom. The global Hamiltonian is the sum of two terms 
$H_{glob}=H_{eff}+H_{add}$. In fact, the existence of resonant laser field
would take influence on the effective Hamiltonian when we take some
approximations. In order to keep the validity of deriving $H_{eff}$, the
coupling between local atom and the laser field must be assumed much weak.
Certainly, this can be easily assured by weakening the amplitude of the
laser field or localizing the atom in the cavity since the coupling rate
depends on the atom position as $\sin (k_{0}z)\exp [-(x^{2}+y^{2})/\omega
_{0}^{2}]$\cite{20}.

We are interested in the two-atom subsystem entanglement which is widely
used in quantum key distribution and quantum encoding. Wootters proposed a
general measurement of two-qubit entanglement which is named as Concurrence%
\cite{21}

\begin{equation}
C=\max \{0,\lambda _{1}-\lambda _{2}-\lambda _{3}-\lambda _{4}\}\text{,}
\end{equation}%
where $\lambda _{i}$ are the none-negative squre roots of the four
eigenvalues of non-Hermitian matrix $\rho \tilde{\rho}$ with $\tilde{\rho}$
defined as $(\sigma _{y}\otimes \sigma _{y})\rho ^{\ast }(\sigma _{y}\otimes
\sigma _{y})$.

We depict the two-atom entanglement situation in Fig. 2-4 in different
parameter spaces ($J_{12}$, $J_{23}$, $J_{31}$, $B$). In Fig. 2, we adopt ($4
$, $4$, $0$, $0.1$) for dotted line and ($4$, $4$, $0.5$, $0.1$) for solid
line, then plot the entanglement versus time. Firstly, we investigate the
impact of next nearest-neighbour (NNN) atom pair on the entanglement of
nearest-neighbour (NN) atom pair. We see from Fig. 2 that, the existence of
NNN coupling damages the maximal entanglement of NN atom pair. Fortunately,
this coupling can be assigned to tend to zero so that it can be neglected.
In doing so, the phase factor $\phi _{21}$ and $\phi _{32}$ is assumed to be
equal, and phase factor $\phi _{31}$ is properly choosed so that $J_{31}$ is
so much small compares with $J_{12}$ and $J_{23}$. This can be achieved by
selecting special optical fiber. Moreover, NNN coupling does not perturb the
period of NN atoms entanglement. So, in utilizing two-atom entanglement,
though the probability of entangling maybe affected under different fiber
devices, the controlling time of maximal entanglement that is much important
in many quantum information experiments\cite{22} should not be changed.

Because of the diversity of fiber path, the phase shift $\phi _{21}$ and $%
\phi _{32}$ could be easily perturbed. In Fig. 3, we investigate the
influence of nonuniform coupling on the NN entanglement. The parameter space
is ($4$, $4.1$, $0$, $0.1$) with dotted line for atoms 1-2 and solid line
for atoms 2-3. Under this circumstance, three points should be stated: the
general two-atom entanglement is reduced compared with that in Fig. 2; there
appears a patial pressure effect for series connection, the NN atoms that
have stronger coupling share larger entanglement; the general entanglement
period is restricted, thus the controlling time for maximal entanglement is
compressed. So, in constructing physical quantum information channel for
network, the fiber used from point (atom-cavity) to point (atom-cavity) is
preferred to be identical even including dissipation on the fiber.

In Fig. 4, we adopt larger $\Gamma $ than in Fig. 2-3 to study the influence
of the interaction between each atom and the local laser field on the NN
atom entanglement. Once the data bus constituted by fiber is contructed, the
information channel will not be easily improved. While, Fig. 4 indicates
that, the alternating of $\Gamma $ remarkably changes the NN atom
entanglement. The period is further depressed, but the amount of
entanglement is much enhanced compares with the former results. Under the
condition of small $\Gamma $, as is pointed before, changing $\Gamma $ is in
fact modulating the entanglement between NN atoms. What we paying attention
to is entangment signals can be manually controlled through alternating $%
\Gamma $. This kind of modulation effects of optical field on atoms or
magnetic field on spins have already been concerned\cite{23}. Certainly, the
synchronization of $\Gamma $s in each local cavity is important for
modulating the entanglement, the related discussion will be involved
elsewhere.

\section{Conclusion}

We have discussed the preparation of effective two-atom entanglement in a
fiber connected triple-atom system. Using input-output theory for lossing
cavity, after tracing over the field variables, we treated the stationary
system as a triple-body Ising problem. The inversion of the particle number
for two atomic levels was thus analogous with the polarization of electron
spin. Local laser fields were applied to generate and madulate the
entanglement of two atoms. The fiber connected atoms is in fact a kind of
data bus for such a simple multi-atom network, the general influence of
which on the two-atom entanglement was studied. The phase shift of the
photons in the fiber has been found to determine the interaction strength
between two Ising-kind sites. In addition, the dissipation of the photon
information through the fiber should be investigated, while, the impact of
dissipation can also be included in the intearaction strength but acts as a
decaying exponential factor of the length of the fiber, in the form of $%
e^{-\nu L}$, with $\nu $ the dissipation per unit length, $L$ the total
length of the fiber. The local laser field plays a key role in modulating
the entanglemnent properties. So, making an opportune choice between fiber
and local laser field is optimal in generating such entanglement. Certainly,
a practically physical quantum network should be constituted by a large
number of atoms\cite{1,24}. Our system can be extended to one-dimension
atomic chain or higher dimension systems. Thus, they may ture out to be a
candidate for constructing quantum information data-bus.

\begin{acknowledgement}
This work is supported by NSFs of China, under Grant No. 10347103, 10305002
and 60472017.
\end{acknowledgement}

Figure Captions:

Fig. 1: Schematic diagram of the supposed system setup. Three two-level
atoms are located in spatially separated microcavities which are connected
via optical fibers. Two of the cavities are single-sided and one of them are
double-sided.

Fig. 2: Entanglement of atom 1 and 2 versus time for A: (dotted line) $%
J_{31}=0$, B: (solid line) J$_{31}$=0.5.

Fig. 3: Entanglement of A: (dotted line) atom 1 and 2 for $J_{12}=4$, B:
(solid line) atom 2 and 3 for $J_{23}=4.1$ versus time.

Fig. 4: Entanglement of atom 1 and 2 versus time for local coupling strength 
$\Gamma =0.2$.

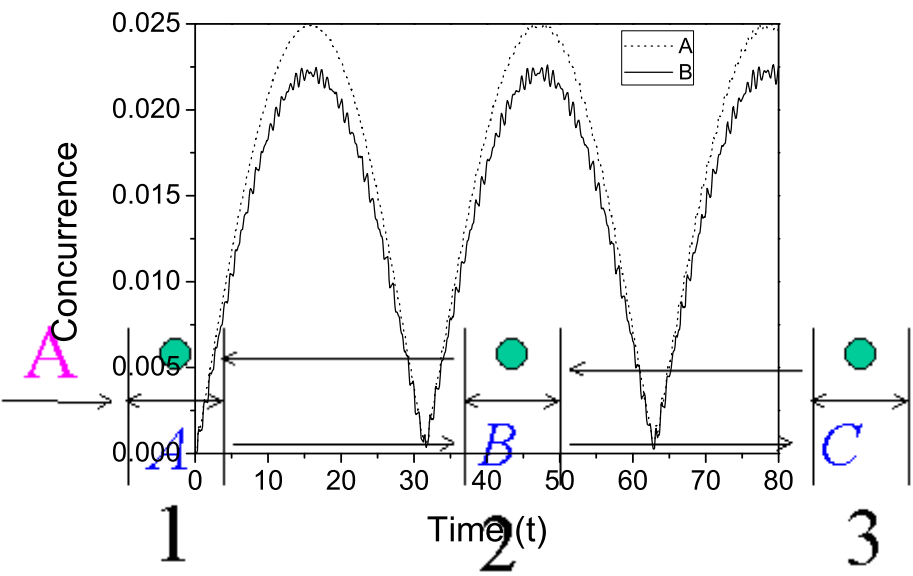
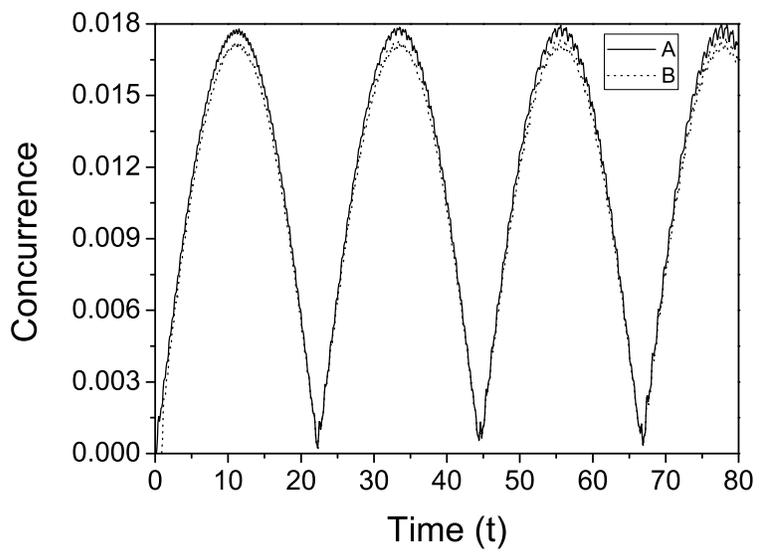
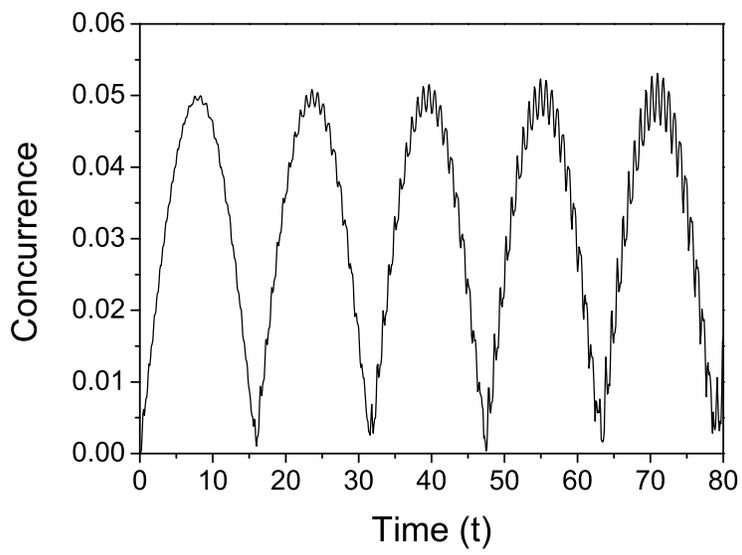
\end{document}